\documentclass{article}
\usepackage{frascatiphys}
\usepackage{graphicx}
\begin{document}
\title{ 
RENORMALON EFFECTS IN TOP-MASS SENSITIVE OBSERVABLES
}
\author{
Silvia Ferrario Ravasio    \\
{\em Institute for Particle Physics Phenomenology, Department of Physics, Durham University, Durham} \\
{\em  DH1 3LE, United Kingdom} 
}


\maketitle
\baselineskip=10pt
\begin{abstract}
A precise determination  of the top mass is one of the key goals of the LHC and future colliders.
Since power corrections are now becoming a source of worry for top-mass measurements,
in these proceedings I discuss the impact of linear infrared renormalons, which plague the definition  of the top pole-mass $m$, on observables expressed in terms of $m$ and in terms of a short-distance mass.
\end{abstract}
\baselineskip=14pt

\section{Introduction}

The top quark is one of the most peculiar particles predicted by the
Standard Model and its phenomenology is entirely driven by the large
value of its mass $m_t$. The most precise measurements of $m_t$ are
based on the use of Monte Carlo~(MC) event generations and the current
errors are of the order of several hundreds of MeV. Thus, linear power
corrections arising from the pole mass ambiguity, which is estimated
to be of the order of 110-250~MeV\cite{Beneke:2016cbu,Hoang:2017btd},
are becoming a major worry in top-mass measurements at hadron
colliders.  Furthermore, even if the perturbative calculations
implemented in the MC generators adopt the pole-mass scheme, there is
still no consensus in the theoretical community regarding the
interpretation of such measurements, due to the complicated interplay
of hadronization and parton shower dynamics\cite{Butenschoen:2016lpz}.
The purpose of these proceedings is not to investigate the relation
between the pole and the MC mass (see
\emph{e.g.}\cite{Hoang:2018zrp}), but instead to investigate the
asymptotic behaviour of quantities calculated in terms of the pole
mass and of the $\overline{\rm MS}$ mass (that we can consider as a
proxy of all the short-distance mass schemes) in a simplified
theoretical frameworks where we understand some aspects concerning the
non perturbative corrections to the pole mass.  We focus upon the case
of single top production and we look at the total cross section, which
is known to be free from physical linear renormalons, the
reconstructed-top mass, which is highly sensitive to the value of
$m_t$, and leptonic observables, which are assumed to be independent
from non-perturbative QCD effects. More details can be found in
Refs.\cite{FerrarioRavasio:2018ubr,FerrarioRavasio:2019glk}.

\section{QCD infrared renormalons}
In gauge theories in general, and in QCD in particular, there is a
certain class of Feynman graphs whose number grows as the factorial of
the order of the perturbative expansion in the strong coupling
constant. The resulting perturbative series is then divergent and it
is typically treated as an asymptotic series. As a consequence, there
is an uncertainty in the value of the sum of the series of the order
$(\Lambda_{\mathrm{QCD}}/Q)^p$, being $Q$ the scale of the process,
$\Lambda_{\mathrm{QCD}}$ the infrared scale at which the validity of
perturbative QCD breaks down and $p$ a positive integer.  This is the
so-called renormalon ambiguity\cite{Beneke:1998ui}.

Indeed, when we perform all-orders calculations, some contributions
can be thought as NLO corrections where the fixed-scale coupling is
replaced with the running one.  After the removal of the UV and IR
divergencies, the perturbative series will take the form
\begin{equation}
  Q^{-p}\int_0^Q d \ell \,\ell^{p-1}\, \alpha_s( \ell) \approx  Q^{-p} \sum_{i=0}^\infty \alpha_s^{n+1}(Q)\int_0^Q \ell^{p-1}d \ell \left(b_0 \log \left(\frac{Q^2}{\ell^2}\right) \right)^n = \sum_{i=0}^\infty\frac{n!}{p}\left(\frac{2 b_0}{p}\right)^n \alpha_s^{n+1}(Q),
\label{eq:asymptotic}
\end{equation}
where $\ell$ is the (real or virtual) gluon momentum, $p$ is a
positive integer and $b_0$ is the one-loop QCD $\beta$ function
\begin{equation}
b_0 = \frac{11 C_A}{12 \pi} - \frac{n_l T_R}{3\pi},
\end{equation}
with $n_l$ being the number of light flavours.  Since $b_0$ is positive,
the series in eq.~\ref{eq:asymptotic} is not even Borel
resummable. The terms in the series will first decrease until
\begin{equation}
  \frac{n!}{p}\left(\frac{2 b_0}{p}\right)^n \approx \frac{(n+1)!}{p}\left(\frac{2 b_0}{p}\right)^{n+1} \alpha_s(Q)\quad \Rightarrow \quad n \approx \frac{p}{2 b_0 \alpha_s(Q)}.
\end{equation}
At this point, if we want to interpret the series as an asymptotic
one, we need to truncate the expansion and the size of the last term,
which is also an indication of the ambiguity in our result, will be of
the order $(\Lambda_{\mathrm{QCD}}/Q)^p$.  The dominant ambiguities
are the ones corresponding to $p=1$, \emph{i.e.} the linear
renormalons, and those affect the definition of the pole mass.

Performing all-order calculations is however not possible for any
non-trivial gauge theory. To overcome this task, we can imagine that
the number of flavours $n_f$ is large and the dominant corrections
arise from $g\to q\bar{q}$ splittings. Thus, everytime we encounter a
gluon line, we replace the free propagator with the dressed one
\begin{equation}
\frac{-i g^{\mu \nu}}{\ell^2+i\eta} \to \frac{-i g^{\mu \nu}}{\ell^2+i\eta}\times \frac{1}{1+\Pi(\ell^2+i\eta, \mu^2)-\Pi_{\rm ct}},
\end{equation}
where $\mu^2$ is the renormalization scale, $\Pi$ is the fermionic
contribution to the vacuum polarization and $\Pi_{\rm ct}$ is the
counterterm we introduce to renormalize the strong coupling. In
$D=4-2\epsilon$ dimensions we can write
\begin{equation}
\Pi(\ell^2+i\eta, \mu^2)-\Pi_{\rm ct} = -  \alpha_s(\mu) \frac{n_f T_R}{3\pi} \left[ \log \frac{|\ell^2|}{\mu^2} - i \pi \theta(\ell^2) +C\right] +\mathcal{O(\epsilon)},
\end{equation}
where $C$ is a renormalization-scheme dependent constant ($C=-5/3$ in
the $\overline{\rm MS}$ scheme). To recover the non-abelian behaviour
of QCD, we can imagine that $n_f$ is large and negative. At the end of
the computation we match the fictitious number of flavours $n_f$ with
the real number of light flavours $n_l$
\begin{equation}
n_f \to n_l - \frac{11 C_A}{4}    = - \frac{3\pi b_0}{T_R}, 
\end{equation}
so that the vacuum polarization appearing in the dressed gluon
propagator takes the desired form
\begin{equation}
\Pi(\ell^2+i\eta, \mu^2)-\Pi_{\rm ct} =   \alpha_s(\mu) \,b_0\, \left[ \log \frac{|\ell^2|}{\mu^2} - i \pi \theta(\ell^2) +C\right]+\mathcal{O(\epsilon)}.
\label{eq:vacpol}
\end{equation}
This is the so-called large-$b_0$
approximation\cite{Beneke:1994qe,Ball:1995ni}.

\section{Single-top production at all orders }

\begin{figure}[tb]
\includegraphics[width=0.24\textwidth]{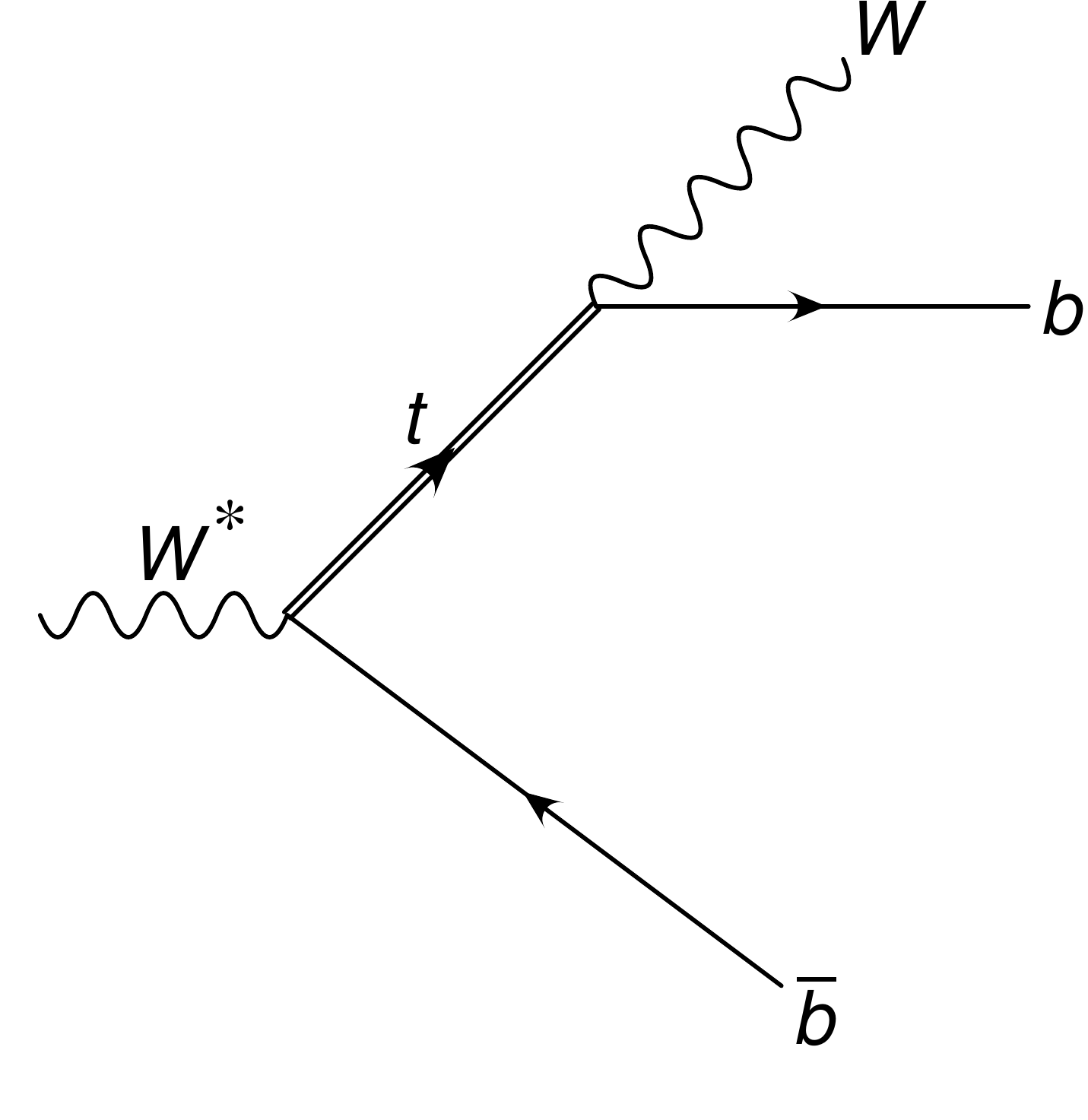}
\includegraphics[width=0.24\textwidth]{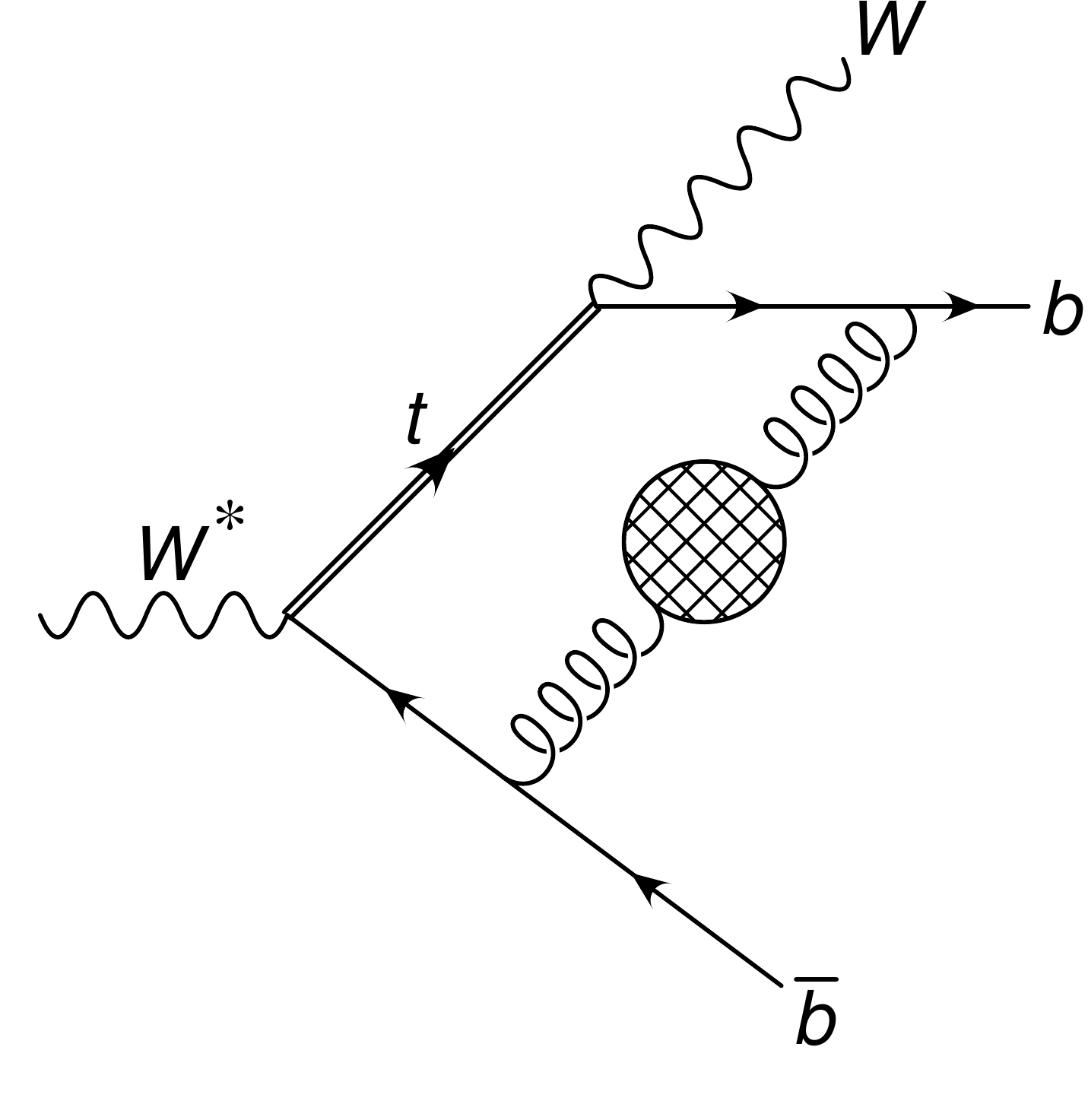}
\includegraphics[width=0.24\textwidth]{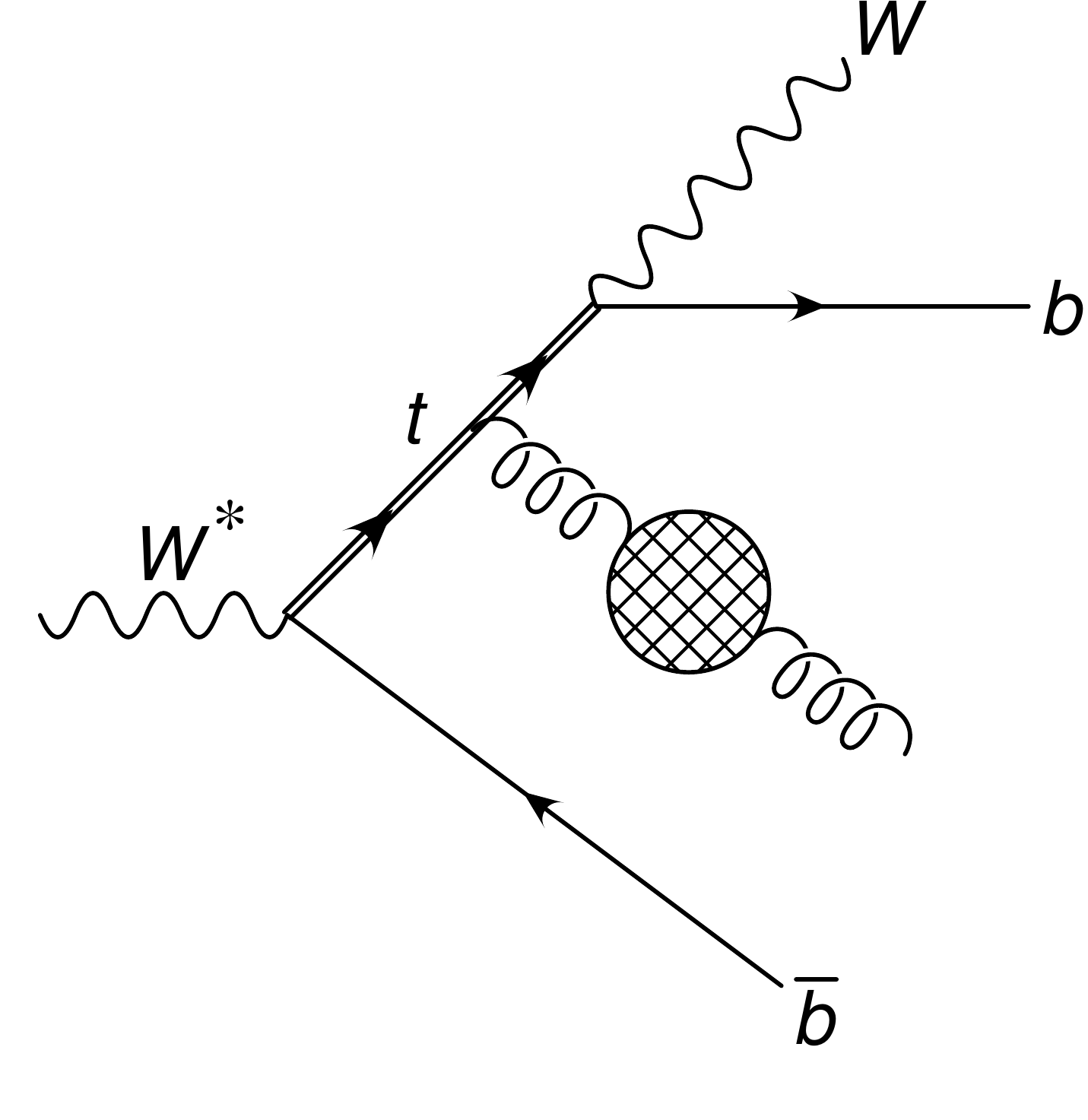}
\includegraphics[width=0.24\textwidth]{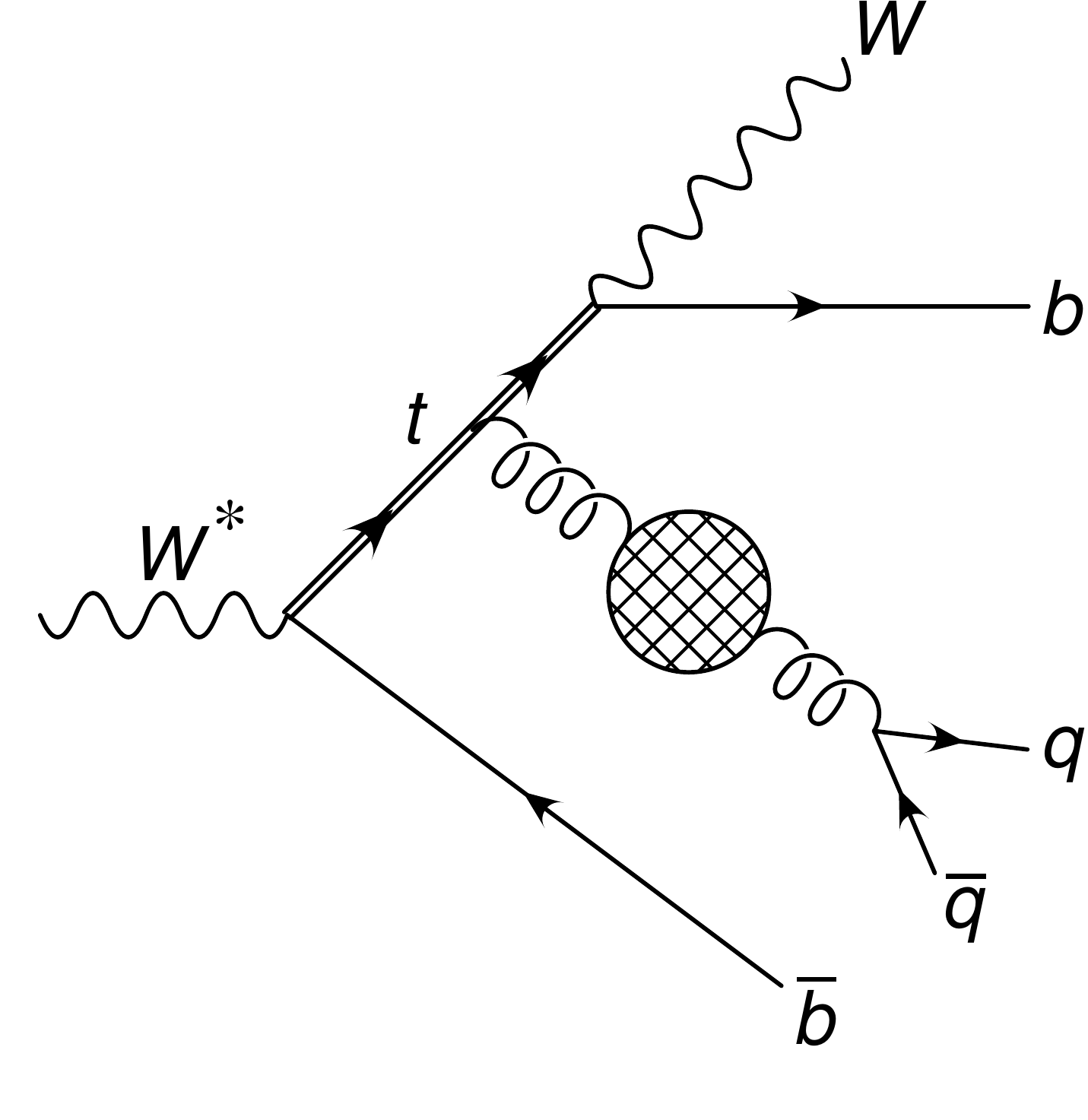}
\caption{Feynman diagram for Born $W^* \to W b \bar{b}$ process, and
  samples of Feynman diagrams for the virtual and the real-emission
  contributions and for the $W^* \to W b \bar{b} q \bar{q}$
  production. The bubble denotes the insertion of the vacuum
  polarization of eq.~(\ref{eq:vacpol}) in the gluon propagator.}
\label{fig:diagr}
\end{figure}

We now calculate the process of single-top production and decay, $W^*
\to t \bar{b} \to W b \bar{b}$, at all-orders in the large-$b_0$
approximation.  Explicative examples of the diagrams that must be
considered are illustrated in Fig.~\ref{fig:diagr}. We stress that
together with the virtual and real corrections where the gluon line
has been dressed, we also need to include the contribution arising
from a real $g \to q \bar{q}$ splitting.

The expression for the total-cross section\footnote{We can obtain the
  expression of the average value of an observable $O$ from the one of
  the total cross-section replacing $\Theta(\Phi)$ with
  $\frac{\Theta(\Phi)}{\sigma^{(0)}}\left[ O(\Phi) - \langle O
    \rangle^{(0)} \right]$ in $T(\lambda)$, where $\langle O
  \rangle^{(0)}$ is the Born prediction.}  in presence of selection
cuts (that we denote with $\Theta(\Phi)$, being $\Phi$ a phase space
point) is given by
\begin{equation}
  \sigma = \int d \Phi \frac{d \sigma}{d \Phi}\Theta(\Phi) = \sigma^{(0)} -\frac{1}{\pi b_0} \int_0^\infty d \lambda \frac{d}{d\lambda}\left[\frac{T(\lambda)}{\alpha_s(\mu)}\right] \arctan\left[\pi b_0 \alpha_s( \lambda e^{-C/2}) \right] 
\label{eq:sigmaall}
\end{equation}
where $\sigma^{(0)}$ is the Born cross section, $C$ is the
renormalization-scheme dependent constant that we choose in such a way
that
\begin{equation}
  \alpha_s( \lambda e^{-C/2}) =  \alpha_s( \lambda) + \alpha_s^2( \lambda) \,b_0\, C  +\mathcal{O}(\alpha_s^3) \equiv  \alpha_s( \lambda) + \frac{\alpha_s^2( \lambda)}{2\pi}\left[\left(\frac{67}{18}-\frac{\pi^2}{6} \right) C_A -\frac{5}{2} n_l \right] = \alpha^{\rm CMW}_s( \lambda), 
\end{equation}
where CMW denotes the Catani-Marchesini-Webber renormalization scheme
for the strong coupling\cite{Catani:1990rr}, also known as the Monte
Carlo scheme. The function $T(\lambda)$ is given by
\begin{equation}
  T(\lambda) = \sigma^{(1)}(\lambda) + \frac{3 \lambda^2}{2 T_R \alpha_s(\mu)} \int d \Phi_{g^*} d \Phi_{\rm dec} \frac{d \sigma^{(2)}_{q\bar{q}}(\Phi)}{d\Phi} \left[ \Theta(\Phi) - \Theta(\Phi_{g^*}) \right],
\end{equation}
where $\sigma^{(1)}(\lambda)$ is the $\mathcal{O}(\alpha_s)$ cross
section calculated with a gluon of mass $\lambda$,
$\sigma^{(2)}_{q\bar{q}}$ is the leading-order cross section for the
process $W^* \to W b \bar{b} q \bar{q}$, $\Phi_{g^*}$ is the
phase-space for the production of a heavy gluon of mass $\lambda$,
$\Phi_{\rm dec}$ the phase-space for its decay into a $q\bar{q}$ pair
(so that the total phase space $\Phi$ can be written as $d\Phi =
\frac{d \lambda^2}{2\pi} d \Phi_{g^*} d \Phi_{\rm dec}$). Thus we see
that the factor $T(\lambda)-\sigma^{(1)}(\lambda)$ takes into account
the fact that the event in which the $q\bar{q}$ pair has been
clustered in a massive gluon $g^*$ can lead to different kinematics
with respect to the full event.  This term is closely related to the
Milan factor\cite{Dokshitzer:1997iz}.

It is easy to check that the $\mathcal{O}(\alpha_s)$ expansion of
eq.~\ref{eq:sigmaall} is given by $\sigma^{(0)}+\sigma^{(1)}(0)$, as
expected.  From eq.~\ref{eq:sigmaall} we also see that we have a
linear renormalon if
\begin{equation}
  \frac{d T(\lambda)}{d \lambda} \Big|_{\lambda=0} \neq 0,
\end{equation}
so we will focus our attention on the small-$\lambda$ behaviour of the
function $T(\lambda)$ to assess the presence of linear renormalons.

\section{Results}
In this section we present the most relevant phenomenological results
of Ref.\cite{FerrarioRavasio:2018ubr}. The center-of-mass energy is
chosen to be $E=300$~GeV, the $W$ mass is set to 80.4~GeV and the
bottom mass is set to 0.  We choose the complex pole scheme for a
consistent treatment of top-offshell effect
\begin{equation}
m^2 = m^2_0 - i m_0 \Gamma_t, 
\end{equation}
where $m_0=172.5$~GeV, $\Gamma_t=1.3279$~GeV. We choose $m_0$ as
renormalization scale.  We use the $e^+e^-$ version of the anti-$k_T$
algorithm to reconstruct the $b$ and $\bar{b}$ jets. If not specified,
we require the $b$ and the $\bar{b}$ jets to be separated and to have
a minimum transverse momentum of 25~GeV.

\subsection{Cross section}

\begin{figure}[tb!]
\includegraphics[height=0.28\textheight]{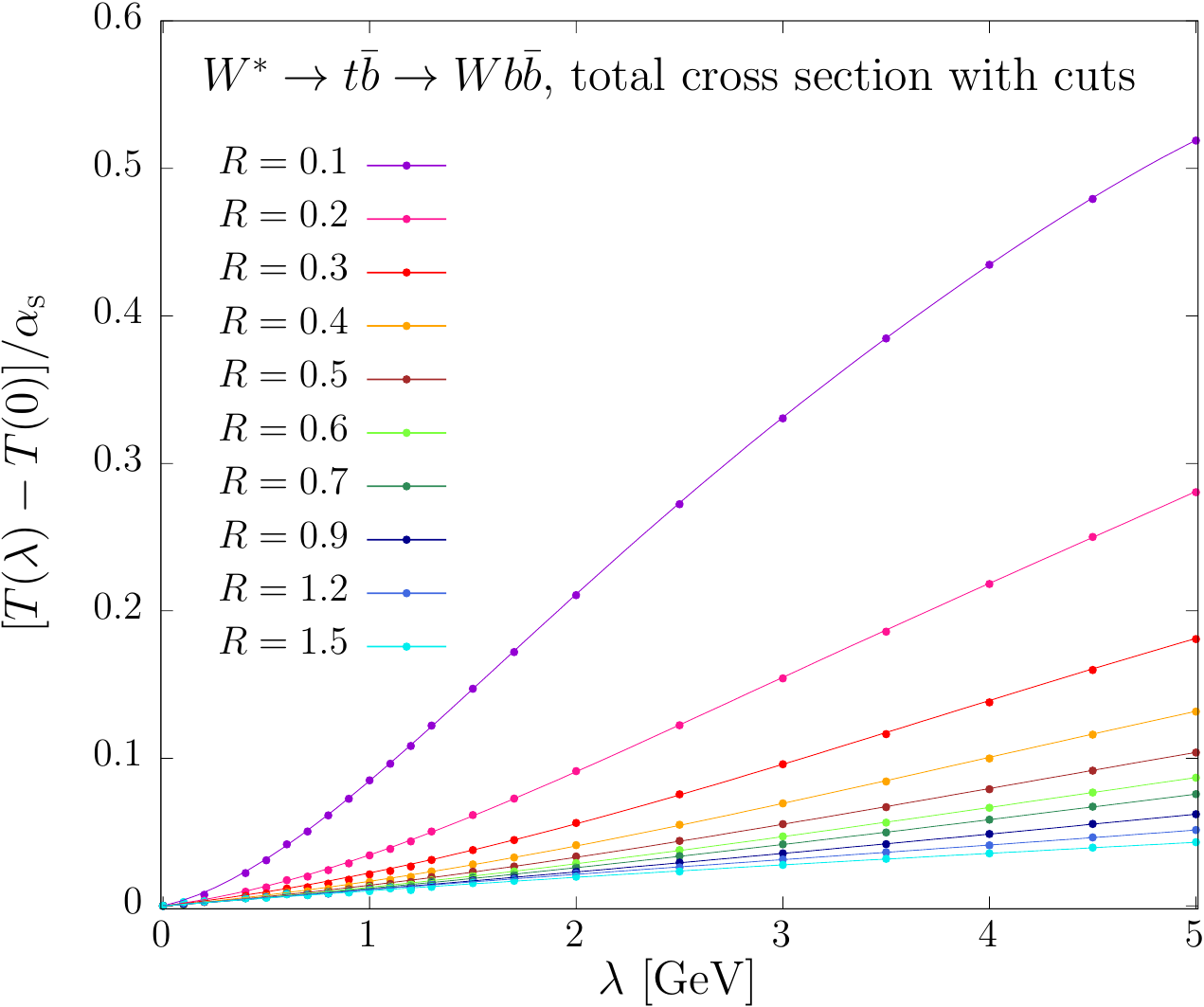}\qquad
\includegraphics[height=0.28\textheight]{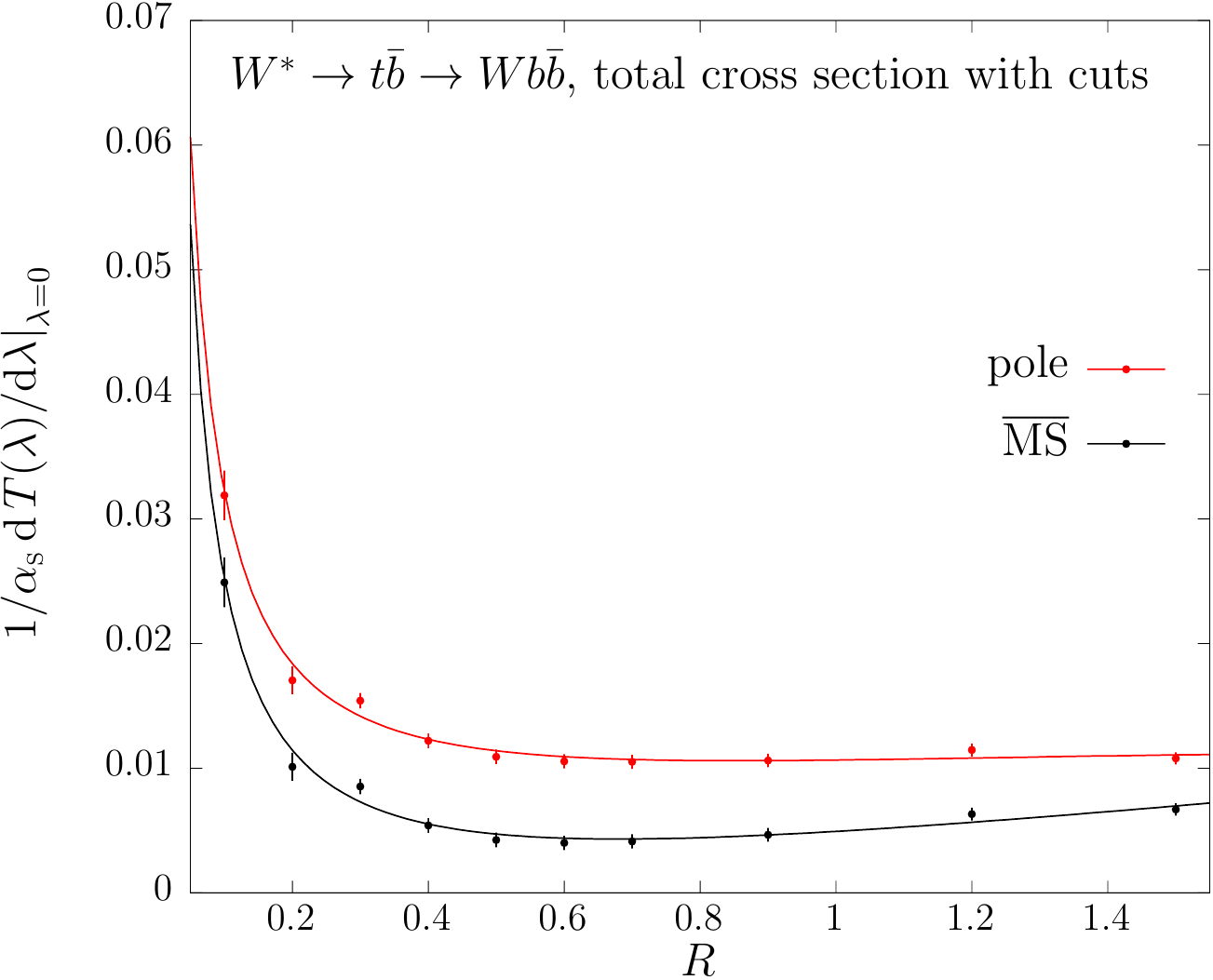}
\caption{In the left pane the small-$\lambda$ behaviour for
  $T(\lambda)$ for the total cross section with cuts calculated in the
  pole scheme for several jet radii. In the right panel the slope of
  $T(\lambda)$ at $\lambda=0$ for the pole and the $\overline{\rm MS}$
  scheme. }
\label{fig:xsec}
\end{figure}

For the total cross section without cuts the function $T(\lambda)$
reduces to $\sigma^{(1)}(\lambda)$. For small values of $\lambda$, the
linear $\lambda$ term is due to the pole-mass counterterm and is equal
to
\begin{equation}
  \frac{d T(\lambda)}{d \lambda} \Big|_{\lambda=0} = \alpha_s(\mu) \frac{C_F}{2} \frac{\partial \sigma^{(0)}(m,m^*)}{\partial {\rm Re}(m) },
\end{equation}
where ${\rm Re}(m)$ denotes the real part of the top mass.  By
expanding eq.~(\ref{eq:sigmaall}) in series of $\alpha_s(\mu)$, we
find that the minimal term is reached at the $8^{\rm th}$ order and
leads an ambiguity of relative order $5\times10^{-4}$.

When the $\overline{\rm MS}$ scheme is employed, such linear
renormalon disappears and the behaviour of the perturbative series
improves, no visible minimum arises considering the first $10^{\rm
  th}$ orders and the relative corrections are smaller then $10^{\rm
  -5}$ already from the $4^{\rm th}$ order.

However, when selections cuts to identify the final state are
introduced, the benefit of using the $\overline{\rm MS}$ scheme is
reduced. The requirement that the $b$ and the $\bar{b}$ jets are
separated and have a minumum transverse momentum of 25~GeV introduces
a linear term whose magnitude grows with the inverse of the jet
radius, as was found in other contexts as
well\cite{Korchemsky:1994is,Dasgupta:2007wa}.  This behaviour is
illustrated in Fig~\ref{fig:xsec}.

\subsection{Reconstructed-top mass}

\begin{figure}[tb!]
\includegraphics[height=0.275\textheight]{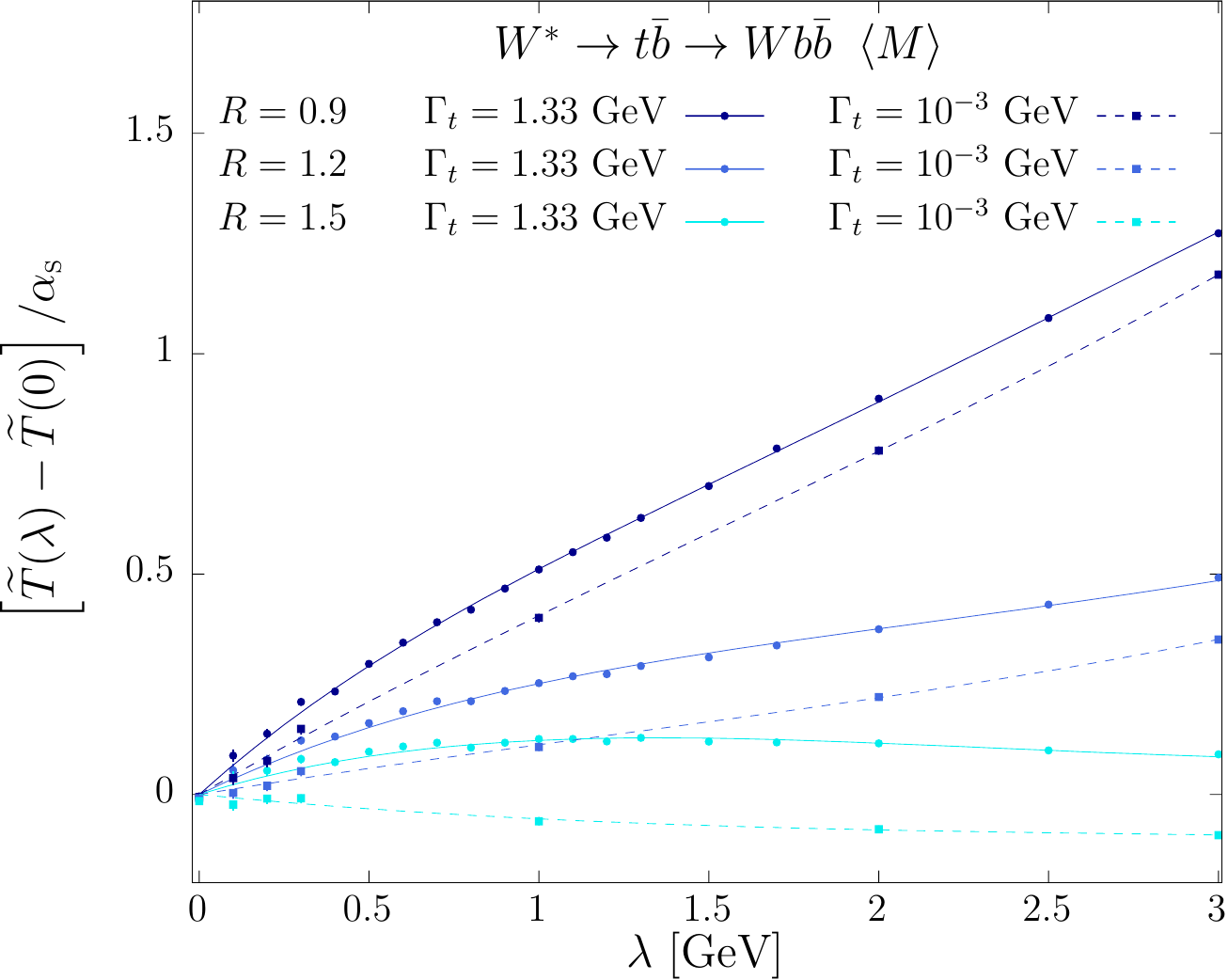}\quad
\includegraphics[height=0.275\textheight]{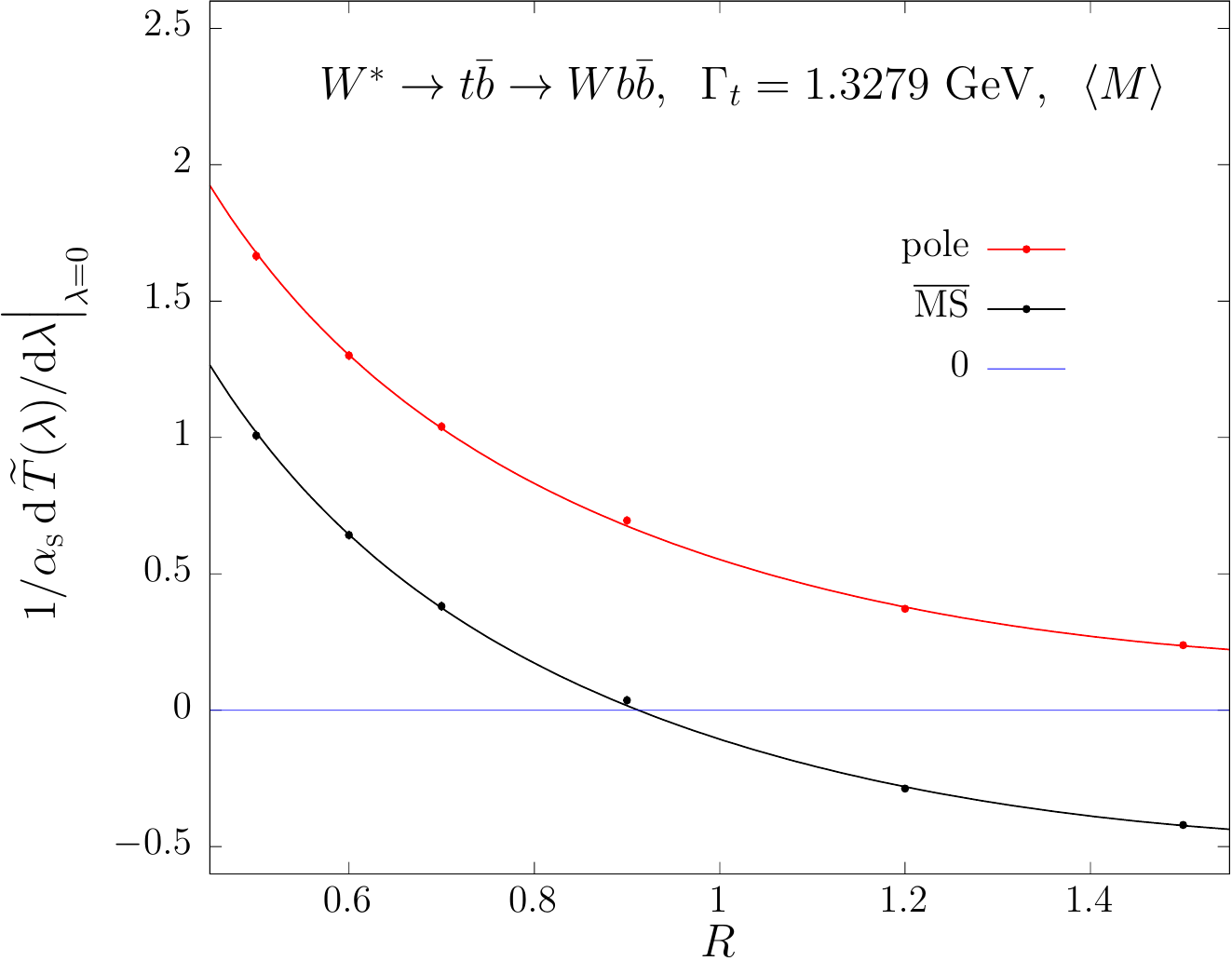}
\caption{In the left pane the small-$\lambda$ behaviour for
  $T(\lambda)$ for the reconstructed-top mass calculated in the pole
  scheme for several jet radii using $\Gamma_t=1.3279$~GeV~(solid
  lines) and $\Gamma_t=10^{-3}$~GeV~(dashed lines). In the right panel
  the slope of $T(\lambda)$ at $\lambda=0$ for the pole and the
  $\overline{\rm MS}$ scheme.}
\label{fig:M}
\end{figure}

We define the reconstructed-top mass $M$ as the mass of the system
comprising the final-state $W$ boson and the $b$-jet.  As for the case
of the cross section, selection cuts introduce a linear-$\lambda$ term
in the function $T(\lambda)$, whose magnitude is proportional to the
inverse of the jet radius.

For vanishing top width, $M$ approaches the pole mass when a large jet
radius is adopted, thus reducing the renormalon ambuiguity. On the
other hand, the use of a short distance scheme like the $\overline{\rm
  MS}$ would introduce a term of the form
\begin{equation}
\frac{1}{\alpha_s(\mu)}\frac{d T(\lambda)}{d \lambda} \Big|_{\lambda=0} = -\frac{C_F}{2} \frac{\partial M(m,m^*)}{\partial {\rm Re}(m) } \approx -\frac{C_F}{2} =-0.667,
\end{equation}
and thus have a worse perturbative expansion.  This behaviour is due
to the fact that this observable contains a physical renormalon that
cancels with the pole renormalon if the pole scheme is adopted.

The inclusion of finite-width effects slightly modifies the slope of
the function $T(\lambda)$ in the range $\lambda < \Gamma_t$, as can be
seen from the left panel of Fig.~\ref{fig:M}. In the right panel of
the same figure we see that for large jet radii there is still a large
cancellation between the physical renormalon present in the definition
of $M$ and the one in the pole mass.  In the $\overline{\rm MS}$
scheme we do observe a cancellation between the jet renormalon and the
one in $M$ for jet radii of the order of 0.9. However, conversely to
the previous case, this cancellation is accidental and cannot be taken
as indication of a small overall ambiguity as the two effects should
be considered independent source of errors.

\subsection{Leptonic observables}

The last observable we consider is the average value of the energy of
the final-state $W$ boson, $\langle E_W \rangle$, which can be
considered as a proxy of all leptonic observables. For this analysis
we do not impose any selection cuts to avoid to be contaminated by jet
renormalons.

We find that in the narrow-width approximation, $\langle E_W \rangle$
has a linear renormalon both in the pole and in the ${\rm
  \overline{MS}}$ scheme.  Conversely to the case of the total cross
section, if we compute $E_W$ in the laboratory frame the calculation
cannot be factorized between production and decay, thus spoiling the
cancellation of the linear $\lambda$ term in $\langle E_W
\rangle$. This cancellation takes place only if $E_W$ is computed in
the top frame.

When a finite width is employed, the top can never be on-shell as
$p_t^2$ is real, thus a linear $\lambda$ term can develop only if the
pole mass counterterm is used. However, this is also telling us that
we can start appreciating the good convergence of the ${\rm
  \overline{MS}}$ scheme at orders $n=1+\log(m/\Gamma_t) \approx 6$,
as it can be seen from Tab.~\ref{tab:EW}.
\begin{table}[h!]
\centering
\begin{tabular}{c|c|c|c|c|}
  \cline{2-5}
  & \multicolumn{4}{|c|}{ $\phantom{\Big|}$ $\displaystyle \langle E_W
    \rangle$ $\phantom{\Big|}$ [GeV]}
 \\ \cline{2-5}
  & \multicolumn{2}{|c|}{ \phantom{\Big|}pole scheme \phantom{\Big|}}& \multicolumn{2}{|c|}{${\overline{\rm MS}}${} scheme}
 \\ \cline{1-5}
 \multicolumn{1}{|c|}{$\phantom{\Big|} i \phantom{\Big|}$}  & $c_i $ & $ c_i\,\alpha_s^i$  & $c_i $ & $ c_i \, \alpha_s^i$ 
 \\ \cline{1-5}
 \multicolumn{1}{|c|}{$\phantom{\Big|}$           1 $\phantom{\Big|}$}& $-1.435 \, (0) \times 10^{    1}$ & $-1.552 \, (0) \times 10^{    0}$ & $-7.192 \, (0) \times 10^{    0}$ & $-7.779 \, (0) \times 10^{   -1}$ 
 \\ \cline{1-5}
 \multicolumn{1}{|c|}{$\phantom{\Big|}$           2 $\phantom{\Big|}$}& $-4.97 \, (4) \times 10^{    1}$ & $-5.82 \, (4) \times 10^{   -1}$ & $-3.88 \, (4) \times 10^{    1}$ & $-4.54 \, (4) \times 10^{   -1}$ 
 \\ \cline{1-5}
 \multicolumn{1}{|c|}{$\phantom{\Big|}$           3 $\phantom{\Big|}$}& $-1.79 \, (5) \times 10^{    2}$ & $-2.26 \, (6) \times 10^{   -1}$ & $-1.45 \, (5) \times 10^{    2}$ & $-1.84 \, (6) \times 10^{   -1}$ 
 \\ \cline{1-5}
 \multicolumn{1}{|c|}{$\phantom{\Big|}$           4 $\phantom{\Big|}$}& $-6.9 \, (4) \times 10^{    2}$ & $-9.4 \, (6) \times 10^{   -2}$ & $-5.7 \, (4) \times 10^{    2}$ & $-7.8 \, (6) \times 10^{   -2}$ 
 \\ \cline{1-5}
 \multicolumn{1}{|c|}{$\phantom{\Big|}$           5 $\phantom{\Big|}$}& $-2.9 \, (3) \times 10^{    3}$ & $-4.4 \, (5) \times 10^{   -2}$ & $-2.4 \, (3) \times 10^{    3}$ & $-3.5 \, (5) \times 10^{   -2}$ 
 \\ \cline{1-5}
 \multicolumn{1}{|c|}{$\phantom{\Big|}$           6 $\phantom{\Big|}$}& $-1.4 \, (3) \times 10^{    4}$ & $-2.2 \, (4) \times 10^{   -2}$ & $-1.0 \, (3) \times 10^{    4}$ & $-1.7 \, (4) \times 10^{   -2}$ 
 \\ \cline{1-5}
 \multicolumn{1}{|c|}{$\phantom{\Big|}$           7 $\phantom{\Big|}$}& $-8 \, (2) \times 10^{    4}$ & $-1.3 \, (4) \times 10^{   -2}$ & $-5 \, (2) \times 10^{    4}$ & $-8 \, (4) \times 10^{   -3}$ 
 \\ \cline{1-5}
 \multicolumn{1}{|c|}{$\phantom{\Big|}$           8 $\phantom{\Big|}$}& $-5 \, (2) \times 10^{    5}$ & $-9 \, (4) \times 10^{   -3}$ & $-2 \, (2) \times 10^{    5}$ & $-4 \, (4) \times 10^{   -3}$ 
 \\ \cline{1-5}
 \multicolumn{1}{|c|}{$\phantom{\Big|}$           9 $\phantom{\Big|}$}& $-3 \, (2) \times 10^{    6}$ & $-7 \, (4) \times 10^{   -3}$ & $-1 \, (2) \times 10^{    6}$ & $-2 \, (4) \times 10^{   -3}$ 
 \\ \cline{1-5}
 \multicolumn{1}{|c|}{$\phantom{\Big|}$          10 $\phantom{\Big|}$}& $-3 \, (2) \times 10^{    7}$ & $-6 \, (5) \times 10^{   -3}$ & $0 \, (2) \times 10^{    6}$ & $-1 \, (5) \times 10^{   -4}$ 
 \\ \cline{1-5}
 \end{tabular}
 \caption{Coefficients of the perturbative expansion of the average $W$-boson energy in the pole and $\overline{\rm MS}$-mass schemes.}
 \label{tab:EW}
\end{table}

The last undesirable feature connected to the use of this observable
is the reduced sensitivity to the top mass. Indeed, for our choice of
the center-of-mass energy $d \langle E_W \rangle / d m\approx0.1$,
while in the top frame $d \langle E_W \rangle / d m\approx 0.4$.

\section{Conclusions}
In these proceedings we have summarized the method introduced in
Ref.\cite{FerrarioRavasio:2018ubr} to evaluate all-orders corrections
in the large-$b_0$ approximation. When the method is applied to
processes involving a decaying top quark, we can predict which
observables are affected by linear renormalons if the pole or a
short-distance mass scheme is adopted. This method is also sensitive
to linear corrections associated with jets.

The total cross section does not display linear renormalons related to
the top mass if a short distance scheme is adopted. This is the case
for leptonic observables only if a finite width $\Gamma_t$ is
employed, unless such observables are computed in the top frame. This
also implies that the good convergence of leptonic-observables
predictions will manifest only at high orders ($n\ge
1+\log(m/\Gamma_t) \approx 6$).  The reconstructed-top mass is
affected by a physical renormalon that partially cancels with the one
contained in the pole mass definition. This cancellation is almost
exact for $\Gamma_t \to 0$ if the jet radius is large enough.

\section{Acknowledgements}
The work summarized here has been carried out in collaboration with
Paolo Nason and Carlo Oleari.  I also want to thank the organisers of
LFC19 for the invitation, particularly Gennaro Corcella, Giancarlo
Ferrera and Francesco Tramontano, the STRONG-2020 network for the
financial support and Tom\'a\v{s} Je\v{z}o for useful comments on the
manuscript.

\end{document}